\documentclass{moriond}

\usepackage{graphicx}
\usepackage{subfig}
\usepackage{amsmath}
\usepackage{amsfonts}
\usepackage{amssymb}
\usepackage{booktabs}
\usepackage{color,slashed}
\usepackage{float}
\usepackage{adjustbox}

\def\be{\begin{equation}}
\def\ee{\end{equation}}
\def\bea{\begin{eqnarray}}
\def\eea{\end{eqnarray}}

\newcommand{\Photo}{}

\begin{document}
\vspace*{4cm}
\title{Probing the Majorana Nature in Radiative Neutrino mass models with the same-sign dilepton final states at future colliders}

\author{ Rachik Soualah }
\address{Department of Applied Physics and Astronomy, \\University of Sharjah, Sharjah, UAE} 
%\address{The International Centre for Theoretical Physics (ICTP), Trieste, Italy.} 
\author{Salah Nasri}
\address{Department of Physics, \\United Arab Emirates University, UAE} 
%\address{The International Centre for Theoretical Physics (ICTP), Trieste, Italy.} 
\author{Safa Naseem}
\address{Department of Applied Physics and Astronomy, \\University of Sharjah, Sharjah, UAE} 

\maketitle\abstracts{
Neutrinos in the Standard Model (SM) are considered to be massless which is in contradiction with the evidence from the neutrino oscillation data. These experiments established that at least two SM neutrinos have non-zero masses and that  individual lepton numbers are violated. This is strong evidence of new physics beyond the SM that should be responsible for generating non zero mass for the neutrinos. 
%\textcolor{red}{I suggest you omit: "The SM light neutrinos are not easy to detect at colliders whereas it is very possible to study the production of heavy neutrinos that can be detected; yielding several signal processes that are categorised according to theirs physics interests"}. 
In this work, we study the collider phenomenology of 
%\textcolor{red}{Don't be very specific; you will discuss the details in the text; I suggest you write instead: 
an  extension  of the SM where neutrinos are generated radiatively  at three-loop.   We show that   the production of  same-sign dilepton   at lepton colliders (such as FCC-ee and ILC) can  be used to  probe the  Majorana nature of neutrinos in this class of models}% by studying the the same-sign dilepton final state at the next generation of large colliders (FCC-ee and ILC) at different high synergies. The focus will be on the same-%sign dilepton production via heavy Majorana neutrinos. It has to be emphasised that this signature is considered very significant discovery channel at the expected lepton collider due to pseudo-free background contributions.  At a %later stage we intend to provide a detailed study/roadmap of the collider phenomenology of neutrino mass models in the hope of further exploring the Dark Matter Phenomenology.}

\section{Introduction}

The Standard Model (SM) of the electroweak and strong interactions is extremely successful in explaining most of the measurements in particle physics at high energies that are accessible today. Nonetheless, despite its enormous success in describing the elementary particle interactions, there appears to be a need for physics beyond the SM at about the
weak scale, with additional features that could explain the presence of Dark Matter (DM) and its nature in the universe ~\cite{dm} in addition to several other questions which remain unanswered like the neutrino masses and mixing ~\cite{nmass}, the origin of the Baryon Asymmetry of the Universe (BAU). 
So far, none of these issues are successfully explained within the SM framework. 
In the SM, the neutrinos are considered to be massless which is in contradiction with the evidence from neutrino oscillation experiments. In addition to this, the absolute neutrino mass scale remains unknown, although it is possible to measure the mass difference between neutrino flavors through neutrino oscillations. Thus, these limitations necessitate the introduction of additional components to the SM and going beyond its actual framework.\\
To tackle this open question, many generic models, which are available in the market, are trying to investigate this issue, for instance: Seesaw mechanism (Type I,II or III)~\cite{seesaw}, Radiative Models: one loop (A. Zee~\cite{Zee}, Ma~\cite{Ma}), two loops (Zee-Babu), or three loops (Krauss, Nasri and Trodden (KNT)). The focus in this work will be on the latter model (KNT) in~\cite{KNT} which proposed the extension of the SM with two electrically charged $SU(2)_{L}$ singlet scalars and one right handed (RH) neutrino field $N_{1}$, where a $\mathbb{Z}_{2}$ symmetry was imposed at the Lagrangian level in order to forbid the Dirac neutrino mass terms. After the breaking of the electroweak symmetry, neutrino masses are generated at three-loop, which makes their masses naturally small due to the high loop suppression. Moreover, the field $N_{1}$ is odd under $\mathbb{Z}_{2}$ symmetry, and thus it is guaranteed to be stable, which makes it a good candidate for DM. It was shown also that in order to fit the neutrino oscillation data and be consistent with different recent experimental constraints such as lepton flavor violation (LFV), one needs to have three RH neutrinos.\\
The aim of this paper is to elaborate an overview as well as systematic assessment of one of the possible signatures for heavy neutrino searches that can be explored mainly at future lepton colliders. 
%Furthermore, for the time being and driven by the overwhelming experimental data available at the ATLAS and CMS experiments at the CERN-Large Hadron Collider (LHC), one could investigate in details any further possible signature at new energy frontiers with its high rate of recorded integrated luminosities at $\sqrt s$=14 TeV and beyond (foreseen in future works).\\
%% 
Within the proposed class of neutrino mass models, the neutrino mass is generated radiatively at three-loop, the lightest right-handed neutrino ($N_{1}$) is a good dark matter candidate, and the scalar sector can give a  strongly first order  electroweak phase transition as required for baryogenesis. In our previous works, we considered similar studies at the CERN Large Hadron Collider (LHC)~\cite{Dil} and the International Linear Collider (ILC)~\cite{ANS}. In this work we focus on new detailed physics studies that will be anticipated on the final states that can be produced due to the new interaction terms and extra matter to the SM Lagrangian introduced by the RH degrees of freedom, which can hint at the existence of the RH Majorana neutrinos. \\

This paper is organised as follows, In Sec. I, we introduce the model and its phenomenology. Then in Sec II, we study the physics analysis and provide the simulation details of the same-sign dilepton signatures  at lepton colliders. Finally, our conclusions and outlook are presented in the last Sec IV.
 
\section{The model and its phenomenology}
%\vspace{0.5cm}
In our work we consider the model ~\cite{KNT}  which is an extension of the SM. It has basically two charged scalar singlets $S_{1,2}$ under $SU(2)_{L}$ gauge group and three right handed neutrinos.  A $\mathbb{Z}_{2}$ symmetry was imposed at the Lagrangian level to forbid Dirac neutrino mass terms. The model Lagrangian is written as: 

\begin{equation}
%\mathcal{L}=\mathcal{L}_{SM}+\{f_{\alpha\beta}L_{\alpha}^{T}Ci\tau_{2}%
\mathcal{L}=\mathcal{L}_{SM}+\{f_{\alpha\beta}L_{\alpha}^{T}Ci\tau_{2}%
L_{\beta}S_{1}^{+}+g_{i\alpha}N_{i}^{C}\ell_{\alpha R}S_{2}^{+}+\tfrac{1}%
{2}M_{N_{i}}N_{i}^{C}N_{i}+h.c\}-V(\Phi,S_{1},S_{2})
\end{equation}

where $f_{\alpha\beta}$ denotes the Yukawa couplings, $\alpha$ and $\beta$ represents the lepton family indices, and $L_{\alpha}$ is the LH lepton doublet. The Majorana RH neutrino masses are $M_{N_{i}}$. The tree-level scalar potential $V(\Phi,S_{1,2})$ is given by:
 \begin{align}
V(\Phi,S_{1,2}) &  =\lambda\left(  \left\vert \Phi\right\vert ^{2}\right)
^{2}-\mu^{2}\left\vert \Phi\right\vert ^{2}+M_{1}^{2}S_{1}^{\ast}S_{1}%
+M_{2}^{2}S_{2}^{\ast}S_{2}+\lambda_{1}S_{1}^{\ast}S_{1}\left\vert
\Phi\right\vert ^{2}+\lambda_{2}S_{2}^{\ast}S_{2}\left\vert \Phi\right\vert
^{2}\nonumber\\
&  +\frac{\eta_{1}}{2}\left(  S_{1}^{\ast}S_{1}\right)  ^{2}+\frac{\eta_{2}%
}{2}\left(  S_{2}^{\ast}S_{2}\right)  ^{2}+\eta_{12}S_{1}^{\ast}S_{1}%
S_{2}^{\ast}S_{2}+\left\{  \lambda_{s}S_{1}S_{1}S_{2}^{\ast}S_{2}^{\ast
}+h.c\right\}
\end{align}

Here $\Phi$ represents the SM Higgs field doublet.
One of the features of this model is that by imposing $\mathbb{Z}_{2}$ symmetry, Dirac neutrino mass term at all levels of perturbation theory, is forbidden. After the electroweak symmetry breaking, the neutrino masses are produced at three-loop, where $N_{1}$ is considered to be the lightest RH neutrino  in the model and hence the most stable and designated candidate for DM.  The three-loop diagram that generates the small non-zero neutrino mass matrix elements is shown in Fig. 1. The model should obey the following constraints:
\begin{enumerate}
  \item The generated small neutrino masses at three-loop fit the neutrino oscillation data.
  \item Furthermore, no conflict with the bounds on Lepton Flavor Violating (LFV) processes, the muon anomalous magnetic moment and $0\nu \beta\beta$ decay.
  \item $N_{1}$ is the considered DM candidate  with a relic density that is in agreement with the observation for masses around the EW scale.
  \item Yields a strong first order electroweak phase transition that ensures any baryogenesis scenario and its testability with the Higgs mass measurement currently provided by the CERN-LHC experiments.
\end{enumerate}

%\textcolor{red}{NO; it is a formula, the  requiring is in the numeric; Omit  "Requiring the most stringent experimental constraints on the main sources of Lepton Flavor Violation (LFV) contribution which arises from"} 
The most stringent bound on lepton flavor violation (LFV) comes from from the processes  $\l_{\alpha} \rightarrow \l_{\beta} \gamma$, with branching fraction given by
% \textcolor{red}{Omit; the branching ratio is then simplified as:} 

\begin{multline}
\mathcal{B}(\mu\to e\gamma)= \frac{\alpha v^4}{384 \pi} \Bigg[
 \frac{ |f_{\mu\tau} f_{\tau e}^{*}|^2}{M_{S_1}^4}
+ \frac{36 |g_{1e} g_{1\mu}^{*}|^2}{M_{S_2}^4}\,F^2_2 (M_{N_1}^2/M_{S_2}^2 )
+ \frac{36 |g_{2e} g_{2\mu}^{*}|^2}{M_{S_2}^4}\,F^2_2 (M_{N_2}^2/M_{S_2}^2 )\\
+ \frac{36 |g_{3e} g_{3\mu}^{*}|^2}{M_{S_2}^4}\,F^2_2 (M_{N_3}^2/M_{S_2}^2 )
 \Bigg]
\end{multline}

%\begin{equation}
%\mathcal{B}(\mu\to e\gamma)= \frac{\alpha v^4}{384 \pi} \left[
% \frac{ |f_{\mu\tau} f_{\tau e}^{*}|^2}{M_{S_1}^4}
%+ \frac{36 |g_{1e} g_{1\mu}^{*}|^2}{M_{S_2}^4}\,F^2_2 (M_{N_1}^2/M_{S_2}^2 )
%+ \frac{36 |g_{2e} g_{2\mu}^{*}|^2}{M_{S_2}^4}\,F^2_2 (M_{N_2}^2/M_{S_2}^2 )
%+ \frac{36 |g_{3e} g_{3\mu}^{*}|^2}{M_{S_2}^4}\,F^2_2 (M_{N_3}^2/M_{S_2}^2 )
%\right ]
%\end{equation}

where $v$ is the vaccum expectation value of $H$, $\alpha$ denotes the the electromagnetic fine structure constant and $F_2$ is the loop function given by $F_2(x)= (1/6(x-1)^4)(1-6x+3x^2+2x^3-6x^2-6x^2\log x)$.
In this case, the considered value shall be 
$\mathcal{B}(\mu\to e\gamma)< 5.7$ x $10^{-13}$ in accordance with the MEG collaboration upper bound ~\cite{MEG}.\\
On the other hand if $\l_{\alpha} = \l_{\beta}=\mu$, new contribution is added to the muon anomalous magnetic moment $\Delta a_\mu$ described as:  

%\begin{equation}
\begin{multline}
\Delta a_\mu = \frac{m_\mu^2}{96 \pi^2} \Bigg[
   \frac{|f_{\mu\tau}|^2 + |f_{\mu e}|^2}{M_{S_1}^2 }
 + \frac{6 |g_{1\mu}|^2}{M_{S_2}^2} F_2( M_{N_1}^2/M_{S_2}^2 )
 + \frac{6 |g_{2\mu}|^2}{M_{S_2}^2} F_2( M_{N_2}^2/M_{S_2}^2 )\\
 + \frac{6 |g_{3\mu}|^2}{M_{S_2}^2} F_2( M_{N_3}^2/M_{S_2}^2 )
 \Bigg]
%\end{equation}
\end{multline}

with the upper limit $\Delta a_\mu <2.89$ x $10^{-9}$~\cite{moment}.

The DM emerges as the lightest RH neutrino in the model ($N_{1}$). The fermion masses and couplings values in the radiative neutrino mass model should fall within the relic density range in~\cite{planck} which represents the abundance of the DM candidate. This reads: 
\begin{equation}
\begin{aligned}
\Omega_{N_{1}}h^{2} & \simeq\frac{1.28\times10^{-2}}{\sum_{\alpha,\beta}|g_{1\alpha}g_{1\beta
}^{\ast}|^{2}}\left(  \frac{M_{N_{1}}}{135~\mathrm{GeV}}\right)  ^{2}%
\frac{\left(  1+M_{S_{2}}^{2}/M_{N_{1}}^{2}\right)  ^{4}}{1+M_{S_{2}}%
^{4}/m_{N_{1}}^{4}} 
& \simeq 0.1199 \pm 0.0027 \text{~\cite{planck}}
\end{aligned}
\end{equation}

% ==== 1
\begin{figure}[t]
\begin{centering}
\includegraphics[width=8cm,height=3.5cm]{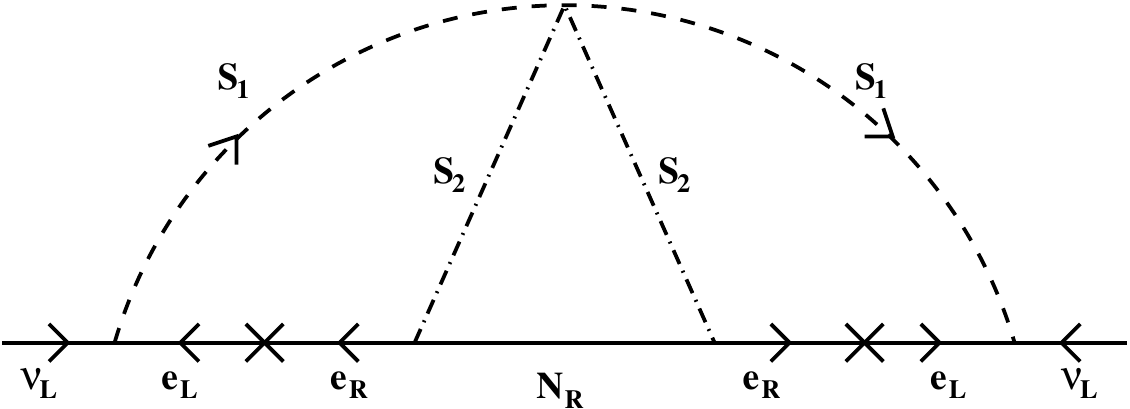}
\par\end{centering}
\caption{The three-loop diagram that generates the neutrino mass.}%
\label{diag}%
\end{figure}

\section{Analysis and discussion}
To validate the model parameters, we check the feasibility of the benchmark that should satisfy all the experimental constraints mentioned above (relic density and lepton flavor violation) and that will be considered at a later stage. In this case, the benchmark selection is explored to determine the range of the model parameters. \\
Considering the benchmark with the model parameter values:

\begin{equation}
\begin{aligned}
f_{e\mu} & =0.02854-i0.03530,~f_{e\tau}=-0.01039+i0.01414, ~f_{\mu\tau}=-0.009136+i0.11839,\nonumber\\
g_{i\alpha} &  = 10^{-1}\times\left(
\begin{array}
[c]{ccc}%
0.0693999-i0.327729 & 0.111002+i0.0806861 & -9.55353+i10.077288\\
-1.72021-i0.83145 & 2.36177+i6.754297 & -7.96116+i10.27103\\
-0.6242649-i4.45564 & -3.37743-i0.8660177 & 6.123282+i0.4577608
\end{array}
\right)  ,\nonumber\\
%\text{and where:} \quad  
m_{N_{i}}  & =\{136.5\mathrm{~GeV},\ 162.1\mathrm{~GeV},~220.2\mathrm{~GeV}%
\},~m_{S_{i}}=\{989.3\mathrm{~GeV},~347.6\mathrm{~GeV}\}
\end{aligned}
\end{equation}

To define the feasible benchmarks, we probe the parameter space of neutrino masses and mixing (after satisfying the LFV, muon anomalous magnetic moment and relic density constraints).
Figure 1 illustrates the range of RH neutrino ($N_{1}$) versus the charged scalars ($S_{1}$ and $S_{2}$) masses that can yield possible DM relic density in the considered parameter phase space (left panel). Clearly the charged scalar masses tend to choose $M_{S_{1}}$ to be larger than $M_{S_{2}}$.  Furthermore, we investigate the Yukawa couplings $g_{i\alpha}$ and $f_{\alpha\beta}$ and their magnitude (right panel). The results are shown on the right side of Figure 1. One can notice that the $f_{\alpha\beta}$  couplings are generally smaller than the couplings $g_{i\alpha}$. Consequently, it becomes mandatory to require the large coupling in the $g_{i\alpha}$ set that satisfy LFV and muon anomalous magnetic moment.\\

\begin{figure}[t]
\begin{centering}
\includegraphics[width=7.9cm,height=5.5cm]{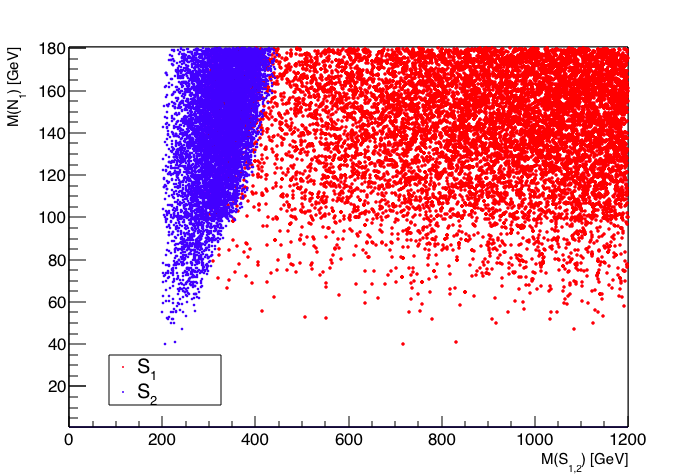}
\includegraphics[width=7.9cm,height=5.5cm]{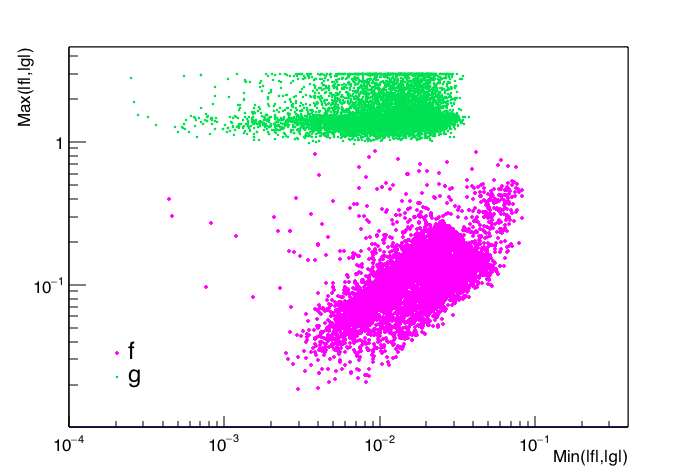}
\par\end{centering}
\caption{Left, the RH Neutrino ($N_{1}$) mass versus the charged scalars ($S_{{1},{2}}$)
masses that can yield possible DM relic density. Right, the Yukawa couplings $g_{i\alpha}$ and $f_{\alpha\beta}$ and their dependance on the maximum and minimum values of their magnitudes.}%
\label{SvsL1}%
\end{figure}
% ========================
% ========================
We study the same-sign dilepton production via heavy Majorana neutrinos at lepton colliders. The natural roadmap would be to initially provide the cross section sensitivity of the same-sign dilepton processes at high energy. In our analysis we take into account the simple setting for neutrino mass spectrum, i.e. $M_{N_{1}} < M_{N_{2}} < M_{N_{3}}$. In this analysis, the considered physics processes are $e^{-}e^{-} \rightarrow e^{-} e^{-} + E_{miss}$ and $e^{-}e^{-} \rightarrow e^{-} \mu^{-} + E_{miss}$. It is worth mentioning that we do not consider the same-sign dimoun final state since it is background free. The missing energy contribution at the final state is $E_{miss} \subset \{ \nu_{e}\bar{\nu}_{e}, \nu_{\mu}\bar{\nu}_{\mu}, \nu_{\mu}\bar{\nu}_{\tau}, \nu_{\tau}\bar{\nu}_{\mu}, \nu_{\tau}\bar{\nu}_{\tau},N_{i}N_{k}\}$ where $i, k = 1, 2, 3$. Figure 2 shows the cross section for each subprocess in $e^{-}e^{-} \rightarrow e^{-} e^{-} + E_{miss}$ (left) and $e^{-}e^{-} \rightarrow e^{-} e^{-} + E_{miss}$ (right) versus the $E_{CM}$ energies. Both possible signature and expected background contributions that can mimic the signal are plotted together.
% ========================
% ========================

\begin{figure}[t]
\begin{centering}
\includegraphics[width=7.9cm,height=6cm]{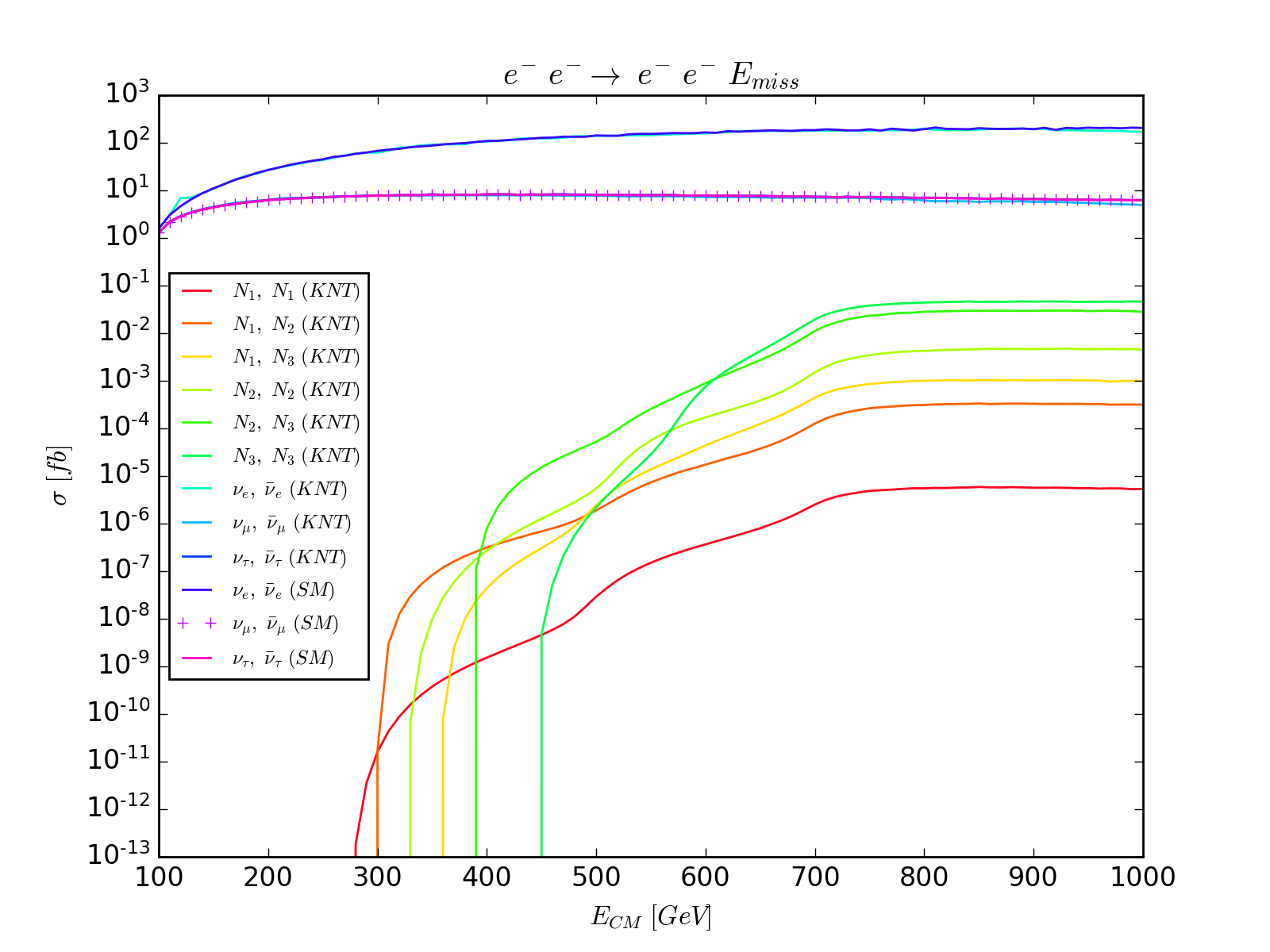}
\includegraphics[width=7.9cm,height=6cm]{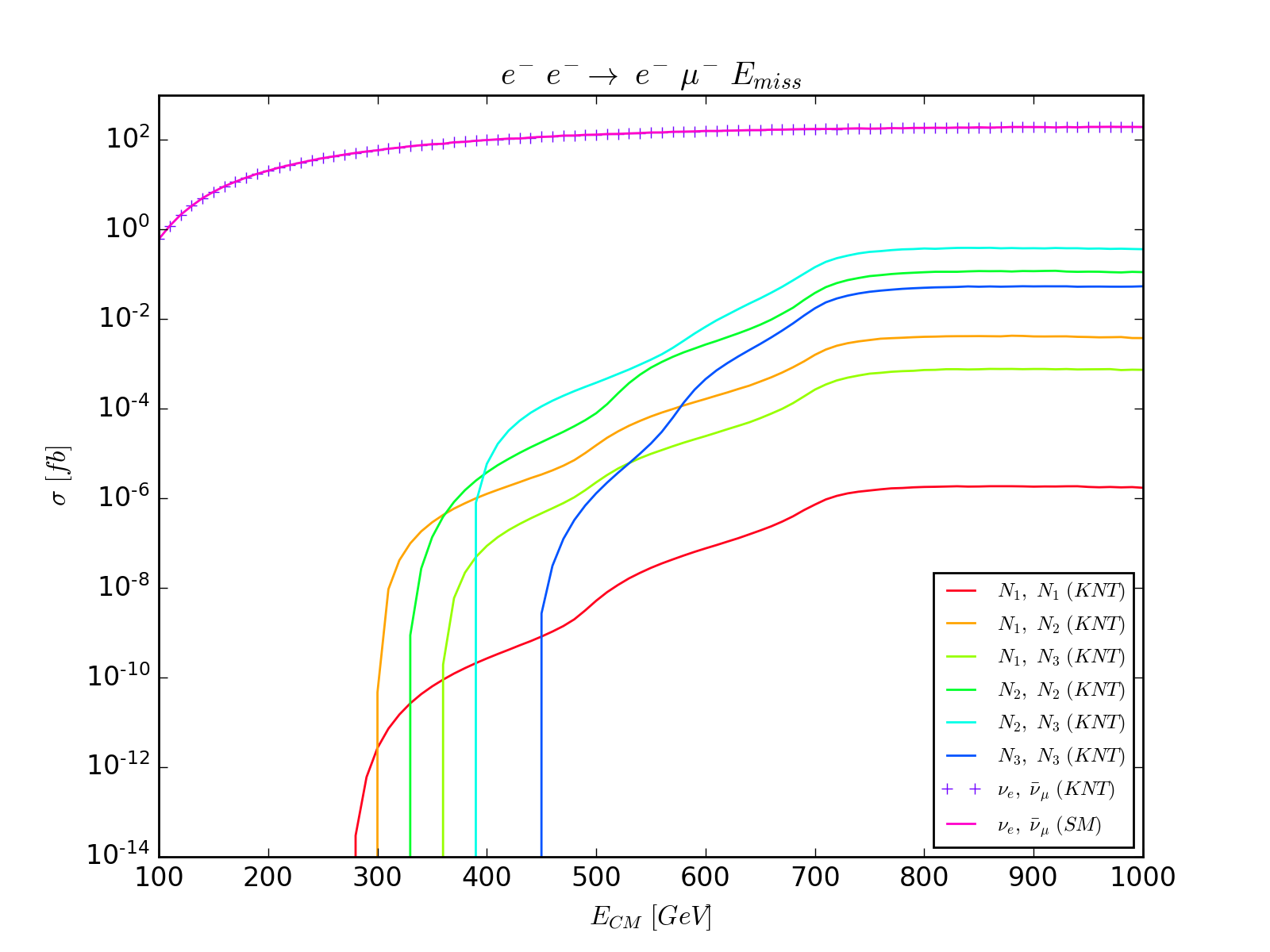}
\par\end{centering}
\caption{The cross section $(\sigma(e^{-}e^{-} \rightarrow l^{-} l^{-} + E_{miss}))$ of the different contributions versus the center of mass energy. The $e^{-}e^{-}$ case is illustrated on the left side whereas the $e^{-}\mu^{-}$ case is shown on the right side.}
\label{SvsL2}%
\end{figure}
% ========================
% ========================
We extensively study the signal and background kinematic distributions in order to define the cut sets where an excess in the number of events in the signal over the background can be observed distinctly. Table 1 summarises the useful kinematic variables and their corresponding cut values in both considered signatures, where an excess in the number of events of the signal over background is prominent. 
We extend our analysis to explore the dependence of the same-sign dilepton production cross section in terms of the mass if the DM candidate ($M_{N_1}$), as well as  the mass of the charged scalar ($M_{S_2}$). The results are shown in Figures 4 and 5 for $M_{N_1}$ and $M_{S_2}$ respectively in both channels at $E_{CM}=1$ TeV.
One can see that all the missing energy contributions can be produced at higher energy whereas at lower energy few of them can be probed due to small cross section values. 

% ========================
% ========================

\begin{table}[!h]
\centering
\begin{tabular}{|c|c|}
\hline
%\multicolumn{2}{|l|}{$e^{-} e^{-} \rightarrow e^{-} e^{-} E_{miss}$}                                                                \\ \hline
Channel & Selection Cuts                                                                                                    \\ \hline
$e^{-} e^{-} \rightarrow e^{-} e^{-} + E_{miss} $            &  \begin{tabular}[c]{@{}c@{}}$120 < E_{e} < 400$, $M_{miss} > 580$,\\ $-0.6 < \cos(\theta_{e}) < 0.8$\end{tabular}  \\ \hline
$e^{-} e^{-} \rightarrow e^{-} \mu^{-} + E_{miss}  $         & \begin{tabular}[c]{@{}c@{}}$20 < E_{e} < 310$, $E_{\mu} > 300$, $M_{miss} > 350$,\\ $-0.8 < \cos(\theta_{e}) < 0.8$, $-0.4 < \cos(\theta_{\mu}) < 0.45$\end{tabular}   \\ \hline
\end{tabular}
\caption{The cut sets for the process  $e^{-}e^{-} \rightarrow e^{-} e^{-} + E_{miss}$ and $e^{-}e^{-} \rightarrow e^{-} \mu^{-} + E_{miss}$ at different CM energies. $E_{e}$ ($E_{\mu}$) represents the energy of the electron(muon), $M_{miss}$ indicates the invariant mass of the possible missing energy contributions. $\theta_{e}$ ($\theta_{\mu}$) refers to the emission angle of the electron (muon) in the final states.
}%
\label{T4}%
\end{table}
%\newpage
% =========
% =========
\begin{figure}[t]
\begin{centering}
\includegraphics[width=7.9cm,height=5.5cm]{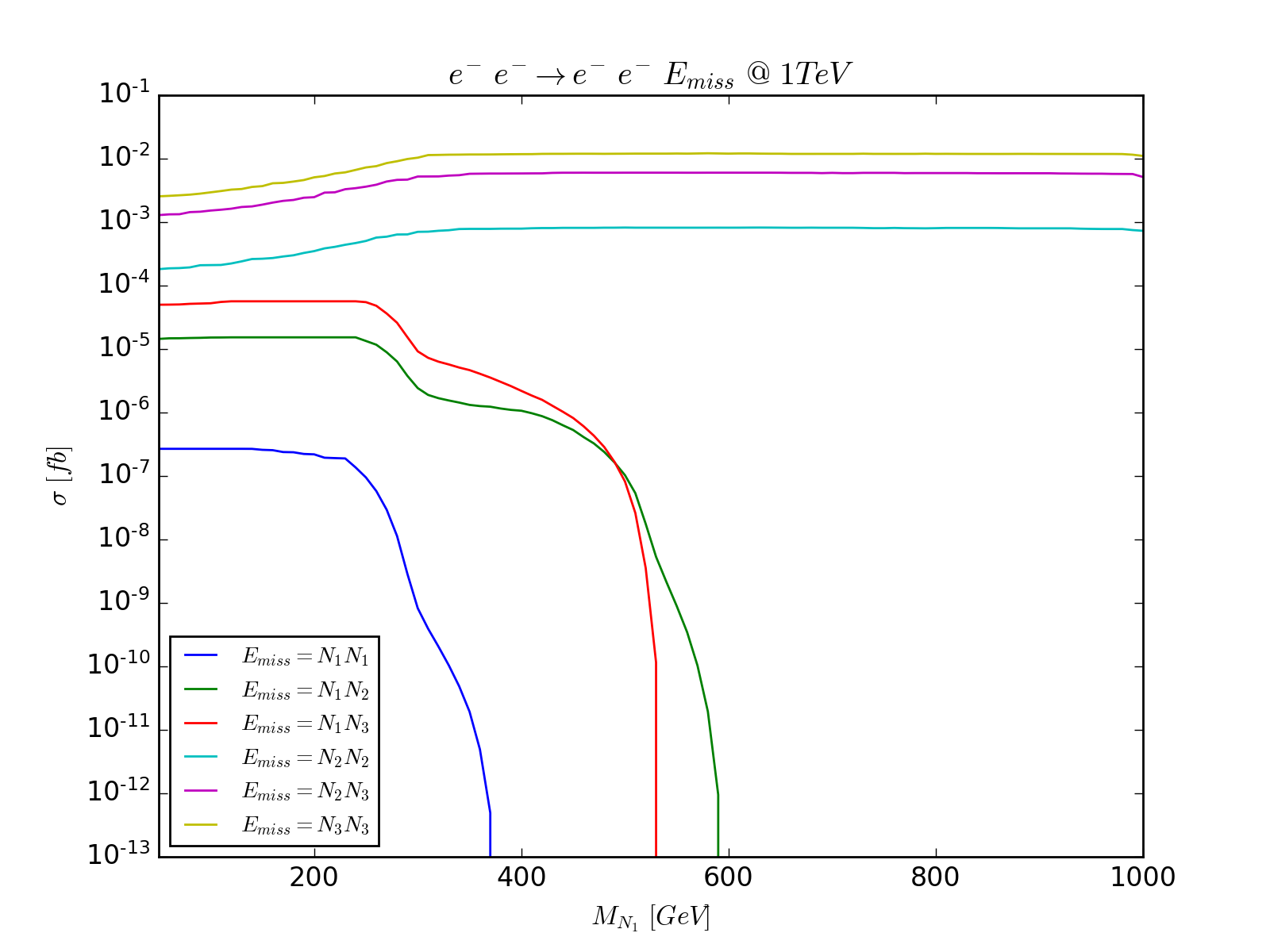}
\includegraphics[width=7.9cm,height=5.5cm]{xs_vs_mn1_em_1000.png}
\par\end{centering}
\caption{The cross section of different subprocesses as a function of the DM candidate mass $M_{N_1}$ for $e^{-}e^{-} \rightarrow e^{-} e^{-} + E_{miss}$ (left) and $e^{-}e^{-} \rightarrow e^{-} \mu^{-} + E_{miss}$ (right) at $E_{CM}=1 TeV.$ }.%
\label{SvsL3}%
\end{figure}

\section{Conclusions}
Although this work is still ongoing, the preliminary results show promising sensitivity of the same-sign dilepton production via Majorana neutrinos in Radiative Neutrino mass models at lepton colliders~\cite{RSS}. The study will be extended to investigate the detectability of a larger class of radiative neutrino mass models at future lepton colliders using polarized beams where the signal is enhanced and can be observed with smaller integrated luminosity as compared to the case of unpolarized beams.

\begin{figure}[t]
\begin{centering}
\includegraphics[width=7.9cm,height=5.5cm]{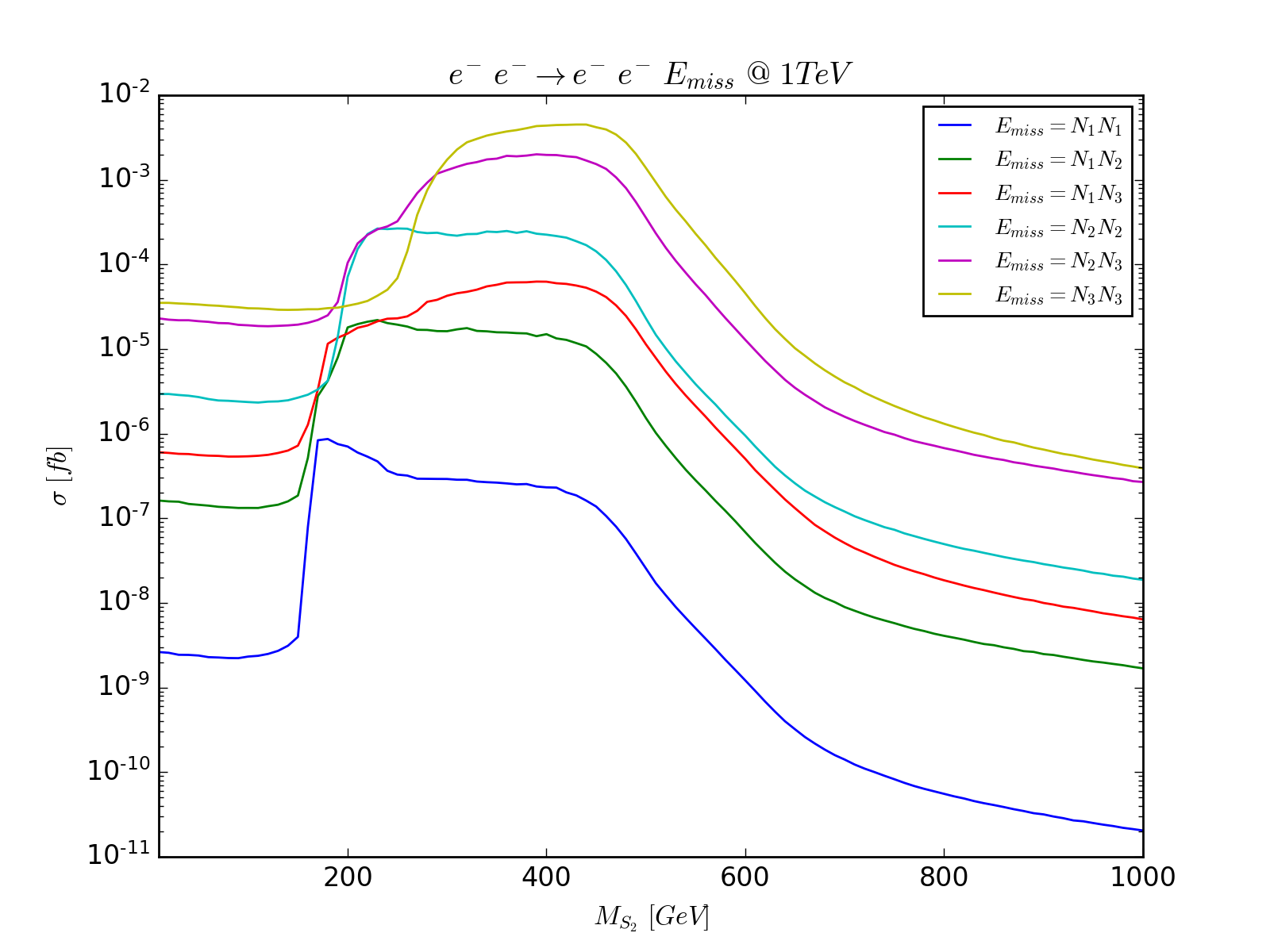}
\includegraphics[width=7.9cm,height=5.5cm]{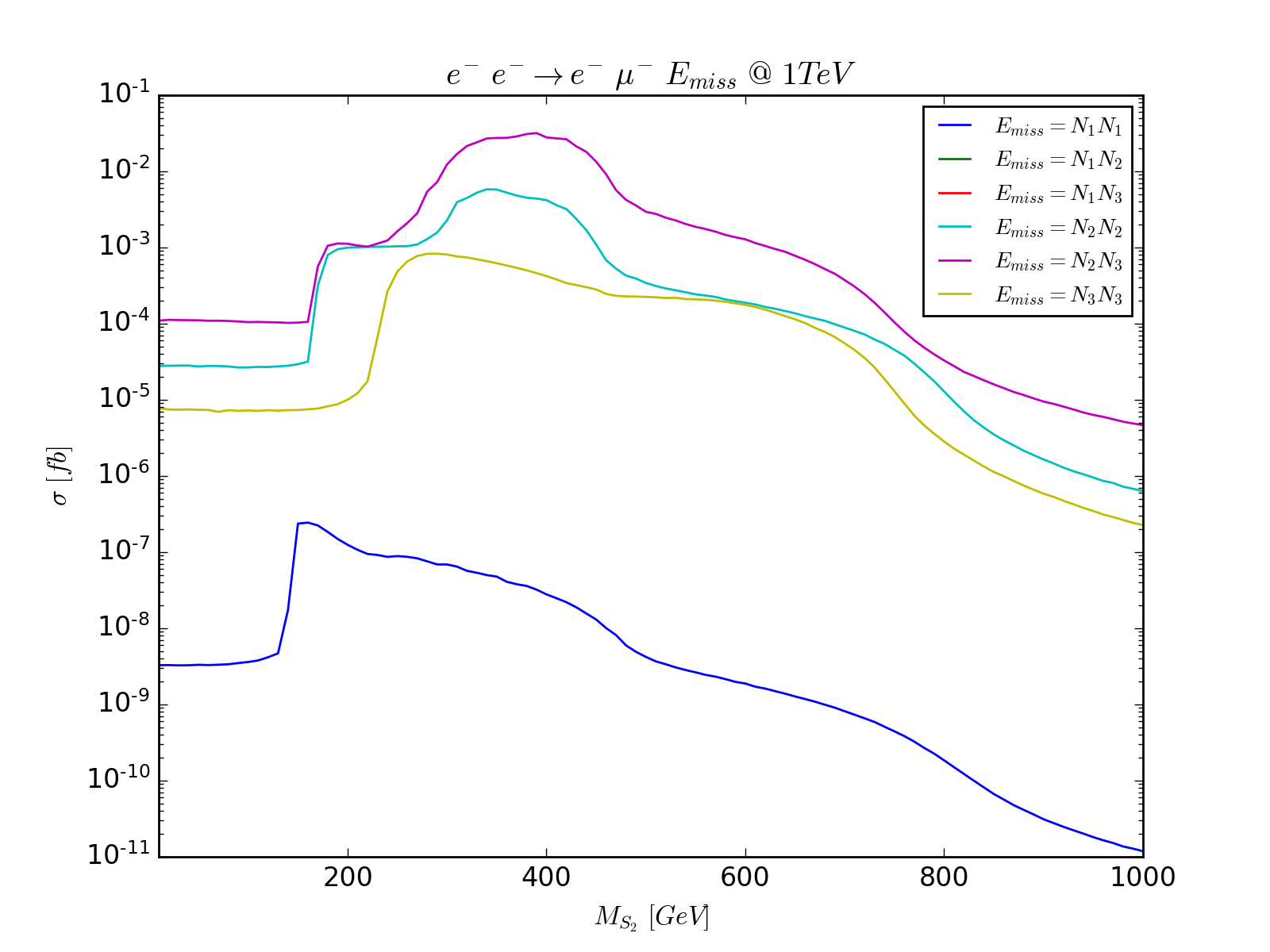}
\par\end{centering}
\caption{The cross section of different subprocesses as a function of the charged scalar mass $M_{S_2}$ for $e^{-}e^{-} \rightarrow e^{-} e^{-} + E_{miss}$ (left) and $e^{-}e^{-} \rightarrow e^{-} \mu^{-} + E_{miss}$ (right) at $E_{CM}=1 TeV.$ }%
\label{SvsL4}%
\end{figure}

\end{document}